\documentclass[pre,twocolumn,showpacs]{revtex4} 
\usepackage{amsmath,amssymb,
graphics,graphicx,epsfig} 
 
\begin{document} 
\newcommand{\BE}{\begin{equation}}  \newcommand{\EE}{\end{equation}} 
\newcommand{\D}{\partial}           \newcommand{\Div}{{\rm{div}}} 
\newcommand{\eps}{\varepsilon}      \newcommand{\taup}{\tau_{ph}} 
\newcommand{\BA}{\begin{eqnarray}}     \newcommand{\EA}{\end{eqnarray}} 
 
\title{Condensation and vortex formation in Bose-gas upon cooling} 
\author{E.A. Brener$^{(a)}$,  S.V. Iordanskiy$^{(b)}$,  R.B. Saptsov$^{(b)}$} 
\affiliation{$^{(a)}$ \it Institut f\"ur Festk\"orperforschung,  
Forschungszentrum J\"ulich, 52425 J\"ulich, Germany \\ 
$^{(b)}$ \it Landau Institute for Theoretical Physics, RAS, 119334 Kosygin str.2, Moscow, 
Russia} 
 
\begin{abstract} 
The mechanism for the transition of a Bose gas to the superfluid state via  
thermal fluctuations is considered. It is shown that in the process  
of external cooling some critical fluctuations (instantons) are formed above 
the critical temperature. The probability of the instanton formation is calculated 
 in the three and two-dimensional  cases. It is found that this 
probability increases as the system approaches the transition temperature. It 
is shown that the evolution of an individual instanton is impossible without the 
formation of vortices in its superfluid part.  
\end{abstract}

\maketitle 
\section{Introduction} 
The ideas of the kinetics of phase transitions have  
been thoroughly developed for first-order phase transitions  
and envisage the existence of the metastable phase  
itself and an equilibrium critical nucleus. The corresponding  
theory was worked out in  \cite{Bec}, \cite{Zel} and described  
in detail in  \cite{Lan}.  
However, theoretical concepts concerning  
the kinetics of second-order phase transitions,  
where these two facts do not exist, have been developed  
insufficiently. In the  \cite{Lif} was 
proposed a certain special  
model for the formation of an ordered phase after the  
fast phase-transition stage in the ''short-range'' order in  
the presence of only two types of ordering. 
 
The interest in the problem of a phase transition  
upon a fast change in external parameters (e.g., temperature)  
has been aroused in connection with the cosmological  
ideas of the Big Bang, where the rapidly  
expanding Universe must be cooled and pass through a  
series of phase transformations accompanied by a  
change in the symmetry of physical fields 
\cite{Zell}.  
It was  
proposed that the kinetics of these transformations can  
be modeled in condensed matter \cite{Kib}. 
 
In the \cite{Zur} was proposed a theory of the second-order  
phase transition upon a rapid change in temperature in  
liquid  $^4 \rm{He}$. The main assumption in the proposed mechanism  
is about the ''critical retardation'' of all processes  
in the vicinity of the transition temperature and ''fast'' 
formation of the nuclei of a new phase upon the subsequent  
cooling. This gives rise to a large number of  
defects on the order of the number of fluctuations far  
above the transition point.

However, no retardation in the formation of a new  
phase has been detected experimentally; the critical  
retardation is associated with the duration of the equilibration  
process at macroscopic distances, which is  
insignificant for the nonuniform process of formation  
of a new phase. 
 
In this work, we consider the transition to a new  
phase via the evolution of fluctuations on scales much  
smaller than the correlation length, which can occur  
quite rapidly even in the vicinity of the critical temperature.  
The transition kinetics in this case is found to be  
directly related to the cooling process itself.  
In our preliminary paper \cite{BIS} 
this approach was proposed for the specific problem of  
Bose condensation of a weakly interacting Bose gas. 
The  appropriate set of equations governing critical  
fluctuations was derived. In the present work a  
detailed analysis of the main equations is performed 
including numerical calculations and  a generalization 
to the case of the two-dimensional exciton gas. 
We will consider the formation of a condensate in the model of  
a weakly nonideal Bose gas with external cooling and  
demonstrate an analogy with  first-order phase  
transitions.

A similar approach to the problem  of the wave nucleation rate 
due to  thermal noise was considered in \cite{Henry}. But the problem of the 
present work requires the noise to be connected to the random thermal fluxes. In 
this case it is appropriate to use a more general  approach for the 
instanton formation namely local Hamilton equations rather than    
the Lagrangian equations 
used in \cite{Henry} (see also \cite{Marder}).  
In \cite{Cher}  the problem of large negative 
gradients in Burgers turbulence was considered,  
which has some resemblance to the differential 
equations discussed in our work, but the boundary problem is quite different. 
 
\section{Dilute Bose-gas upon cooling} 
 
The standard theory of a weakly nonideal Bose gas  
involves a Hamiltonian of the form : 
\BE 
\label{gam} 
\hat{H}=\sum \limits_p \frac {\hat{p}^2} {2m} \hat{a_p}^{+} \hat{{a}_p}+\frac{2\pi \hbar^2 a_0} m \sum \limits_p 
\hat{{a}_{p_4}}^{+} \hat{{a}_{p_3}}^{+} \hat{{a}_{p_2}} \hat{{a}_{p_1}}, 
\EE 
where $a_0$ -- is the scattering amplitude having the atomic  
scale and $m$ -- is the atomic mass.  The properties of such  
gas for a small density $n$ (determined  
by the gas parameter $ \eta=n {a_0}^3 \ll 1$ ) are close to the  
properties of an ideal Bose gas with the transition temperature \cite{Ld}: 
\BE 
\label{tc} 
T_c=\frac {3.31} {\sqrt{2}} \frac{\hbar^2 n^{2/3}}{m}. 
\EE 
At temperatures below the transition point, the ideal  
Bose gas has a pressure depending only on the temperature: 
\BE 
\label{press} 
P_{id}=0.0851 \frac {m^{3/2} T^{5/2}} {\hbar^3}, 
\EE 
which corresponds to zero isothermal sound velocity.  
 
Considering the finite scattering amplitude  
 we can write the qualitative equation of state  
below the transition point as 
\BE 
\label{pressni} 
P=P_{id}(T)+ \frac {\hbar^2 a_0 n^2} m. 
\EE 
We omitted the insignificant constant factor in the second  
term.

The entire kinetics is essentially determined by the  
Bose-gas cooling mechanism. We will consider a simple  
model where the Bose gas is in a certain solid  
matrix with which it only slightly interacts. Such a situation  
may take place, for example, for the exciton gas  
in a crystal. The crystal can be rapidly cooled to a low  
temperature; in this case, the Bose-gas cooling proceeds  
via phonon emission. Assuming that the heat  
capacity of the crystal is large compared to the Bose  
gas, we can disregard the presence of thermal phonons  
in the crystal and their effect on the Bose gas. As a  
result, we obtain a uniform energy-loss mechanism,  
which is described by a phenomenological quantity  
$ 
\label{losses} 
 T/\taup  
$. 
The other models of cooling necessitate the analysis  
of heat transfer at the sample boundaries, which is  
a much more complicated problem. 
The loss rate  $  1/ {\taup} $   is determined by the collisions of particles with each  
other and by the interaction with phonons, which will  
be regarded as weak: 
$$ 
\label{inequal} 
1/\taup \ll 1/\tau_{tr}. 
$$ 
Since  $1/\taup\sim nv_T \sigma_{ph}$ ,  
$1/\tau_{tr}\sim nv_T {a_0}^2$ ($v_T$ s the thermal  
velocity), 
  this means that  
$$ 
\label{crossesction} 
\sigma_{ph} \ll {a_0}^2, 
$$ 
where $\sigma_{ph}$ is the cross section for scattering with  
phonon emission, which corresponds to the weak interaction  
of Bose gas with the crystal. 
 
In view of the smallness of quantity $ 1/ {\taup}$ the evolution  
of the Bose system is slow; in particular, we  
assume that the acoustic wavelength $c \taup \sim v_T \taup $ is  
large compared to the characteristic length  
$\sim \sqrt{\chi\taup}$ , where $ \chi $ - is the thermal diffusivity:    
\BE 
\label{sound} 
\frac {\sqrt{\chi\taup}} {v_T \tau_{ph}}\sim \sqrt{\frac {l^2}{{v_T}^2 
\tau_{tr}\taup}}\sim  
\sqrt{\frac{\tau_{tr}}{\taup}} \ll 1 
\EE 
($l$ stands for the mean free path). 
This makes it possible  
to assume that the fluctuation evolution occurs at a constant  
pressure that coincides with the thermodynamically  
equilibrium pressure.  
 
It follows from Eq.(\ref{pressni}) that the density  
variation $\delta n$ in the fluctuation region is related to a change in temperature  
by  
\BE 
\label{nt} 
\frac {\delta n} n = -\frac {\delta T} T \frac 1 {\eta^{\frac 1 3}}. 
\EE 
The relative density fluctuation is large compared to the  
relative temperature fluctuation in the temperature  
range 
$T<T_c$. 
This leads to a rapid increase in the reciprocal  
phonon time  
\BE 
\delta \frac 1 {\taup}\sim  -\frac {\delta T} T \frac 1 {\eta^{\frac 1 3}} 
\frac 1 {{\taup}^0} 
\EE 
(where $1/ {{\taup}^0}=n v_T \sigma_{ph}$) with decreasing temperature  
and enhancement of cooling in the fluctuation region.  
For this reason, we will disregard the phonon emission  
in the region far from the developed fluctuations,  
assuming that  
\BE 
 \frac 1 {\taup}\sim  -\frac {\delta T} {T_c}  
\frac 1 {\eta^{\frac 1 3}}\frac 1 {{\taup}^0}U(T_c-T), 
\EE 
where $U(T_c-T)=1$ for $\delta T= T-T_c <0$ and $U(T_c-T)=0$ for $T-T_c>0$. 
 
This allows us to consider the problem of fluctuation  
kinetics within the framework of the theory of hydrodynamic  
fluctuations by supplementing the hydrodynamic  
equations with the energy flux carried away as a  
result of phonon emission: 
\BE 
\frac T {{\taup}^0} \frac {\delta n} n = -\frac {T_c-T} {{\taup}}   U(T_c-T), 
\EE 
where $1/{\taup}=(1/{\taup}^0)(1/\eta^{1/3})$. 
In view of the constancy  
of pressure, we can describe the evolution of temperature  
fluctuations by the heat conduction equation 
\BE 
\label{teq} 
n c_p\left( \frac {\D T} {\D t} \!+\! \vec{v} \frac {\D T} {\D 
\vec{r}}\right)\!=\! 
\nabla(\varkappa\nabla T)\!+\! \frac {T-T_c} {{\taup}}  n c_p  U(T_c\!-\!T), 
\EE 
where the energy flux carried away by phonons is  
added, $\varkappa$ -- is the heat conductivity, $c_p$ -- is the specific  
heat per particle under a constant pressure. This equation  
contains the drift term with mass velocity  
$v({\bf r})$, which appears due to the high density in the fluctuation  
core. In the following analysis, this term will be omitted  
as a higher-order term in fluctuation. 
We are interested  
in the temperature-field fluctuations and their time evolution.  
To analyze these fluctuations, we must introduce  
random heat fluxes $\bf q $ \cite{Lan}, \cite{Liff},  
i.e., the Langevin term ${\bf\nabla q} $. 
These fluxes are delta-correlated (i.e., correlated at distances  
and time intervals smaller than the hydrodynamic  
scales). 
In the case considered, this is ensured by  
the fact that the time $\taup $ and distance $\sqrt{\chi{\taup}}$  
 ($\chi$ -- is  
thermal diffusivity) are larger than the microscopic  
characteristics.  
 
The probability  
$ W_t(T(r))$ 
of realizing the given configuration $ T(r) $ fluctuation field at time $t$  
obeys the  
Fokker-Planck  
equation in variational derivatives \cite{Kl} 
\BA 
\label{fp} 
&&\frac \D {\D t}W =-\int \frac \delta {\delta T(r)} \Big[  
\frac {\chi {T_{\infty}}^2}{n c_p}  
{\bf\nabla}^2\frac \delta {\delta T(r)}W  \nonumber\\ 
&&\qquad +\left[\chi{\nabla}^2 T + 
U(T_c-T)\frac {T-T_c}{\taup} \right] W\Big]d^3r. 
\EA 
 In the absence of the interaction with phonons, the stationary  
solution to this equation coincides with the  
result obtained in the thermodynamic theory of fluctuations.  
The quantity  
\BE 
\chi {\bf\nabla}^2 T +U(T_c-T)\frac {T-T_c}{{\taup}}=\frac {\D T} {\D t} 
\EE 
is the temperature-variation rate upon the deviation  
from the mean value $T=T_{\infty} $. 
 
We assume that fluctuations occur at a fixed temperature $T_{\infty}>T_c$. 
Fluctuations with  
$\triangle T=T-T_{\infty} \ll T_{\infty}$  occur  
quite frequently and are characterized by a certain (in  
fact, stationary) spatial distribution that determines the  
value of $W_t( T)$. In view of the normalization, the latter  
quantity gives the number of small fluctuations in a unit  
volume. However, rare large-amplitude fluctuations  
with 
$ T{\sim}T_c-T_{\infty}$, $T<T_c$,  also sometimes occur, initiating  
the effective cooling by phonons, so that the fluctuation  
becomes irreversible and the nucleus of a new  
phase appears.  Our goal is to calculate the probability  
of such fluctuations in a unit volume per unit time. Since they are infrequent and the distribution at small  $T_{\infty}-T$  
is stationary, one can use the method of characteristics  
to determine the exponentially low probability  
of formation of such a nucleus (instanton for the Fokker-Planck equation).  
An important difference from the  
theory of nucleation in the first-order phase transition is  
that the probability of instanton formation in this case  
is determined by the cooling process.  
 
\section{A toy model} 
 
To clarify the situation, let us consider the instanton  
solution in the case of one degree of freedom, for which  
the Fokker-Planck equation has the form  
\BE 
\label{1dfp} 
\frac {\D W} {\D t}=\frac \D {\D x}\left(D\frac {\D W} {\D x}-v W\right), 
\EE 
where $D$ is the constant diffusion coefficient and $v(x)$ is  
the macroscopic variation rate of the quantity $x$ with  
allowance for its relaxation upon the deviation from  
equilibrium and for an external effect (analogue of  
phonon emission). 
 Setting $W=e^S$,  and assuming that the  
moduli of S and its first derivative are large, we obtain,  
to leading terms, the equation   
\BE 
\frac {\D S}{\D t}=D{\left({\frac {\D S} {\D x}} \right)}^2-v\frac{\D S} {\D x} 
-\frac{\D v}{\D x}. 
\EE 
This is the Hamilton-Jacobi equation with the Hamiltonian (${\D S}/{\D x} =p$) 
\BE 
\label{hamilton} 
H({\frac {\D S}{\D x}} ,x)=-Dp^2+pv+\frac{\D v}{\D x}. 
\EE 
The Hamilton equations are the characteristics of this  
equation in partial derivatives, 
\BE 
\label{firstham} 
\frac {dx} {dt} = -2Dp+v, 
\EE 
\BE 
\label{secondham} 
\frac {dp} {dt} = -{\frac {dv} {dx}} p -\frac{d^2 v} {d^2 x}. 
\EE 
The contribution of the velocity divergence to the  
Hamiltonian is significant only in the vicinity of the  
point  $v = 0$. We are interested in the special  
solution that passes through the equilibrium point $p = 0$,  
$v = 0$. In the 1D Fokker-Planck equation, one can eliminate  
the term with a first derivative by substitution; in  
this case, we have an analogy with quantum mechanics  
and can use the well-known results. Nevertheless, we  
will use direct estimates in the vicinity of $v= 0$.

In the Hamilton equation, the energy is conserved. 
 In view of the smallness of the divergence term, this  
gives $ H=-D p^2 +pv =0$, whence $p=v/D$ and 
\BE  
 \label{action} 
 S=-\int\frac{v^2} D dt=\int\limits_0 \limits^{x^{*}}\frac{v} D dx.  
\EE  
We assume that the velocity $v(x)$ is a convex-down  
function with two zeros (stable at zero and unstable at  
 $x^*$ ($x^* >0$). Such a shape of the function $v(x)$ is ensured by the  
entire cooling process, including phonon emission. 
For  
$x > x^*$, the solution tends to larger values of x, while the  
action is gathered from zero to $x^*$, where $v< 0$. In the  
vicinity of $x^*$, we must take into account the quantity $ {dv}/{dx}$ . For 
large values of $x$, $p^2$ can be ignored, yielding $p\approx- { ({dv} /{dx})}/ v$,   
\BE 
\label{action2} 
S\sim S_0 - \ln(v/v_0), 
\EE 
where $v_0$ is the effective velocity in the region where  
the solutions for $x<x^{*}$ and $x>x^{*}$ match. The solution  
$S_0$ 
itself has the form $e^S\approx {v_0} e^{S_0} / v$,   
and current $j\approx v_0e^{S_0}$.  
One can estimate the value of $v_0$, assuming that all  
terms in the Hamiltonian $H$ are of the same order of  
magnitude:  
\BE  
 \label{1estim} 
 Dp^2\sim vp\sim \frac {d v}{d x}\sim\frac{v_{max}}{x^{*}} 
\EE 
which gives  
\BE 
\label{2estim} 
v_0\sim \sqrt{\frac{D|v_{max}|}{x^{*}}}\sim \frac{|v_{max}|}{\sqrt{|S_0|}}. 
\EE 
 
In the many-dimensional case, the situation is the  
same,  
\BE 
\frac {dx^i} {dt} = -2D^{ij}p_j+v^i, 
\EE 
\BE 
\frac {dp_i} {dt} = -{\frac {\D v^k} {\D x^i}} p_k -\frac{\D( \Div \vec{v})} {\D x^i}, 
\EE 
where $p = 0$ at the beginning and $p\rightarrow 0 $  at the end of  
the trajectory. Consequently, $|p|$ reaches its maximal  
value somewhere on the trajectory. At this point, the  
matrix ${\partial v_i}/{\partial x_k}$ has one zero eigenvalue and ${\bf p}$ is tangent  
to the corresponding eigenvector; subsequently, the trajectory  
passes to the neighborhood of the point corresponding  
to zero velocity $v$. This leads to the definition  
of the critical fluctuation (instanton) as a solution passing  
through the point ${\bf x} = {\bf p} = 0$, whereupon ${\bf p}\rightarrow 0$ for  
$|x|\rightarrow \infty$ as $1/v$, retaining the probability flux at a constant  
level.  
 
\section{Optimal fluctuation} 
 
An analogous procedure can be carried out for the  
field as well. In this case, the Hamiltonian has the form,  
in accordance with Eq.(\ref{fp}),  
\BE 
\label{ham} 
H\!=\!\int  p({\bf r}) \left[ \frac {\chi {T_{\infty}}^2}{n c_p}  
{\bf \nabla}^2 p({\bf r})\!+\!\chi{\bf \nabla}^2 T \!+\! 
U(T_c-T)\frac {T-T_c}{{\taup}}\right] d^3r 
\EE 
with the Hamilton equations  
\BE 
\label{eqT} 
\frac {\D T} {\D t}= \frac {2\chi {T_{\infty}}^2}{n c_p}  
{\bf \nabla}^2 p({\bf r})+\chi{\bf \nabla}^2 T 
+U(T_c-T)\frac {T-T_c}{\taup}, 
\EE 
\BE 
\label{eqp} 
\frac {\D p} {\D t}=-\chi{\bf \nabla}^2p-\frac {p}{\taup}U(T_c-T). 
\EE 
Here, $p={\delta S}/{\delta T({\bf r})}$. Eqs. (\ref{eqT}), (\ref{eqp}) define the  
critical fluctuation and can be reduced to dimensionless  
variables by the substitutions $\xi=r/\sqrt{\chi\taup}$, $\tau=t/ \tau_{ph}$,  
$$ \Theta=\frac {T-T_c} {T_{\infty}-T_c} , ~~ 
p=\frac{n c_p(T_{\infty}-T_c)\Pi}{{{T}^2}_{\infty}},$$  
where $\Theta$ and $\Pi$  are the new dimensionless fields. In this  
case, the dimensionless equations have the form  
\BE 
\label{et} 
\frac {\partial\Theta} {\partial \tau}={\bf \nabla}^2  
\Theta +\Theta U(-\Theta)+2{\bf \nabla}^2\Pi, 
\EE 
\BE 
\label{ep} 
\frac {\partial\Pi} {\partial \tau}=-{\bf \nabla}^2 \Pi-\Pi U(-\Theta). 
\EE 
The solution should fulfill the conditions  
$\Pi_{\xi \rightarrow\infty} \rightarrow 0 $,  
$\Pi_{\tau \rightarrow -\infty}\rightarrow 0 $, $\Theta_{\tau \rightarrow 
-\infty} \rightarrow 1 $, $\Theta_{\xi \rightarrow 
\infty} \rightarrow 1 $ 
 and pass through the neighborhood of 
${\partial\Theta}/ {\partial\tau }\approx 0 $, $\Pi\approx0 $ at 
$\tau\rightarrow \tau^* $. Later the fluctuation is developed by cooling,  
while the random fluxes can be neglected and ${\partial\Theta}/  
{\partial \tau}\approx{\bf\nabla}^2 \Theta +\Theta U(-\Theta) $. 
With  exponential precision we can assume that 
 ${\partial \Theta}/ {\partial \tau} \rightarrow 0$ at 
 $\tau \rightarrow + \infty$, and $\Theta$ tends to the stationary solution  
 $\Theta_{st}$ of the thermal diffusion equation, Eq.(\ref{et}): 
 \BA 
\label{statsol} 
&&\Theta_{st}=\sin (\xi)/ \xi, ~\xi<\pi~, \nonumber \\ 
&&\Theta_{st}=1-\pi/\xi, ~\xi>\pi~. 
\EA 
 
The difficulties with the numerical solution of  this boundary value problem   
are due to the instability of Eq.(\ref{ep}) for 
ascending time whereas Eq.(\ref{et}) is unstable for descending time. Therefore, it 
is impossible to find numerically the solution of the Cauchy problem in either 
direction of time. We briefly describe our adopted procedure. 
At  early stages of the evolution, when $\Theta > 0$ everywhere in space,   
it is easy to check the validity of the relation 
\BE 
\label{sol} 
\Theta=1-\Pi,   
\EE 
which coincides with the thermodynamical theory of temperature fluctuations. 
The function $\Pi$  grows with time according to Eq. (\ref{ep}) (thermal 
diffusion equation with negative time derivative). We can assume that at $\tau=0$ 
the maximum of  $\Pi$ will reach 1. After this   Eq. (\ref{sol}) is no 
longer valid.  
We can consider Eq. (\ref{ep}) as a Schroedinger equation with  imaginary time and 
$\Pi \rightarrow 0$ at 
 infinite time.  
At large $\tau$, the function $\Theta$ will be close to the stationary solution,  
Eq.(\ref{statsol}), which 
becomes zero  at $\xi=\pi$. 
This means that $\Pi$, at large $\tau$,  has an asymptotic proportional to 
\BE 
\label{asym} 
\Pi_{inf}=\exp{(-|\lambda| \tau)\Psi_{\lambda}},  
\EE 
where $\Psi_{\lambda}$ corresponds to the eigenfunction with the negative eigenvalue, 
$\lambda=-0.4576$, according to the Schroedinger 
equation with the potential $-U(\pi-\xi)$ (all other states will grow with  
$\tau$). 
 
Let us denote by $r^* (\tau) $  the space-point for which $\Theta=0$ at time $\tau$. 
The curve $r^* (\tau) $ is a function of $\tau$ starting  
at small $\tau$ as  square root of 
$\tau$ (because  $\Theta$ has a minimum at $\xi=0$) and 
tending exponentially to $\pi$ at large $\tau$ (according to 
the Schroedinger-equation analogy, Eq.(\ref{asym})). 
We don't know the exact form of $r^* (\tau)$ but we can choose some probe function with 
 the same asymptotic behavior. Having such a probe function, we can numerically 
 solve Eq.(\ref{ep}) integrating it backward 
in time (it is unstable while integrating it forward in time) with the condition 
$\Pi_{inf}=\alpha \Psi_{\lambda}$ (of form Eq. (\ref{asym}))  
(the prefactor $\alpha$ should be 
chosen to satisfy Eq. (\ref{sol}) at $\tau=0$). 
 
Using this function $\Pi$, we can numerically solve Eq. (\ref{et}) in  
two space-time regions independently. The 
first one is the internal region $\xi< r^* (\tau) $ the other one is $\xi> r^* (\tau)$ 
(the external region). 
$ 
\Theta \rightarrow 1 
$ 
for $\xi \rightarrow \infty$ (for the external region)  
and $\Theta (r^* )=0$ (for the both 
regions). 
For the exact function $r^* $ the space derivative $\Theta^{'} ={\partial 
\Theta}/ {\partial \xi} $ will be continuous.  
For our probe function $r^*$  there will be some jump 
of the space derivative $\Theta^{'} $ at $r^* (\tau)$. Thus, after performing the 
calculations, we will have in general a nonzero 
jump-function  
$$ 
\Delta \Theta^{'} = \Theta^{'} (r^*-0)-\Theta^{'} (r^*+0) 
$$ 
depending on the choice of the curve $r^* $. Afterwords we should search  
for a more 
exact $r^*$ in order to minimize $max|\Delta \Theta^{'}|$. Some steps within the 
framework of this procedure have been performed. 
 
We use a space-time grid 2000x2000. The grid spacing was taken as  
$ 
\delta\xi=0.015 
$ 
for the space coordinate and 
$ 
\delta\tau=0.01 
$ 
for the time coordinate. 
First, we perform calculations backward in  time for $\Pi$ using  
a standard implicit numerical 
scheme where the space derivatives are calculated for the final time of each time 
step. There are some modifications of the space grid in the vicinity of $r=r^*(t)$, for 
a  better finite difference representation of the Laplacian. 
 Then we use the  analogous scheme for Eq.(\ref{et})  
and find $\Theta$ in the internal and external regions integrating forward  
in  time and 
obtain $\Delta \Theta^{'}$. 
We have defined our probe function $r^* (t)$ by a number of parameters. 
To obtain an exact solution we need, of course, an infinite number of parameters.  
In practice, for a reasonable precision we  only need a few.  
The simplest  form which  obeys the asymptotic behavior is 
\BA \nonumber 
&&r^* = \alpha \sqrt{\tau} + \beta \tau , ~ \tau < \tau_0, \nonumber \\  
&&r^*= \pi-\delta \exp{(-|\lambda| \tau)}, ~ \tau> \tau_0. \nonumber 
\EA 
The parameters should be chosen such that $r^*$  
is continuous and smooth at $\tau=\tau_0$.  
This means  that we have two free parameters  in this case (e.g. $\alpha$ and $\tau_0$).  
After minimization  of $max |\Delta \Theta^{'} |$ with respect to these two parameters 
we find right and left derivatives 
$\Theta^{'} (r^*\pm 0)$   
(see Fiq.1). 
\hskip 10pt 
 
\begin{figure} 
{ 
\centering\includegraphics[height=6cm]{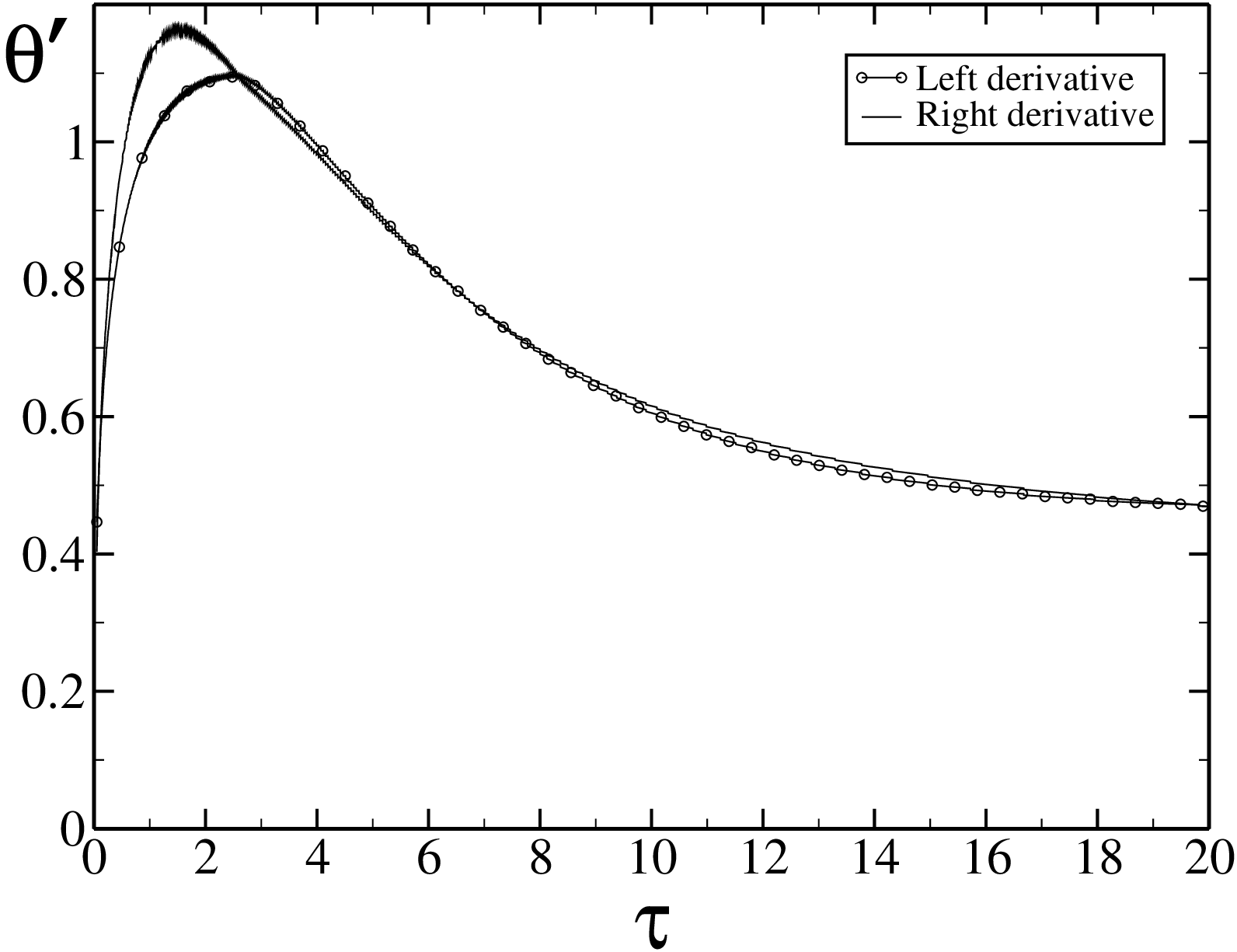} 
\caption{Left and right derivatives at $r^*$ for two free parameters} 
} 
\end{figure} 
 
We can improve  our results  adding a new  term $\gamma\tau^2$:  
\BA 
\label{rgood} 
r^* = \alpha \sqrt{\tau} + \beta \tau + \gamma \tau^2 , ~ \tau < \tau_0, 
\nonumber \\ 
r^*= \pi-\delta \exp{(-|\lambda| \tau)}, ~ \tau > \tau_0.  
\EA 
We have found that for $\alpha=2.28$, $\gamma=0.016$, $\tau_0=2.55$ there is 
an acceptable minimum of $|\Delta \Theta^{'}|$ (see Fig.2).  
The corresponding jump-functions 
 $\Delta \Theta^{'}$ for the two and three parameter cases (for comparison) 
 are plotted in Fig.3. One can see that the inclusion of this additional  term  
 decreases the maximum deviation,  
$max [ |\Delta \Theta^{'}|]$, by a factor of 3. 
\hskip 10pt 
\begin{figure} 
{ 
\includegraphics[height=6cm]{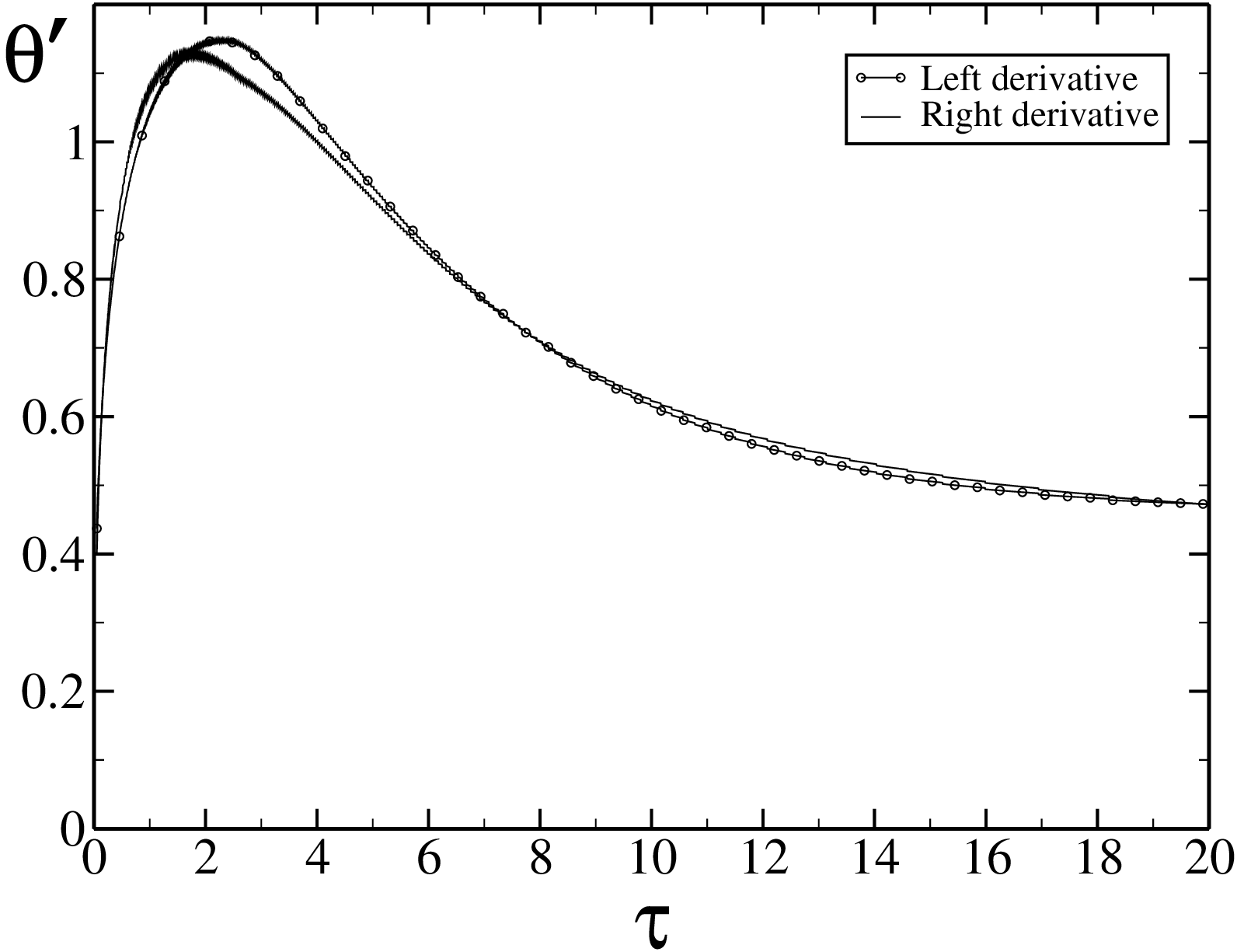} 
\caption{Left and right derivatives at $ r^* $ for three free parameters} 
} 
\end{figure}

\hskip 1pt 
\begin{figure} 
{ 
\includegraphics[height=6cm]{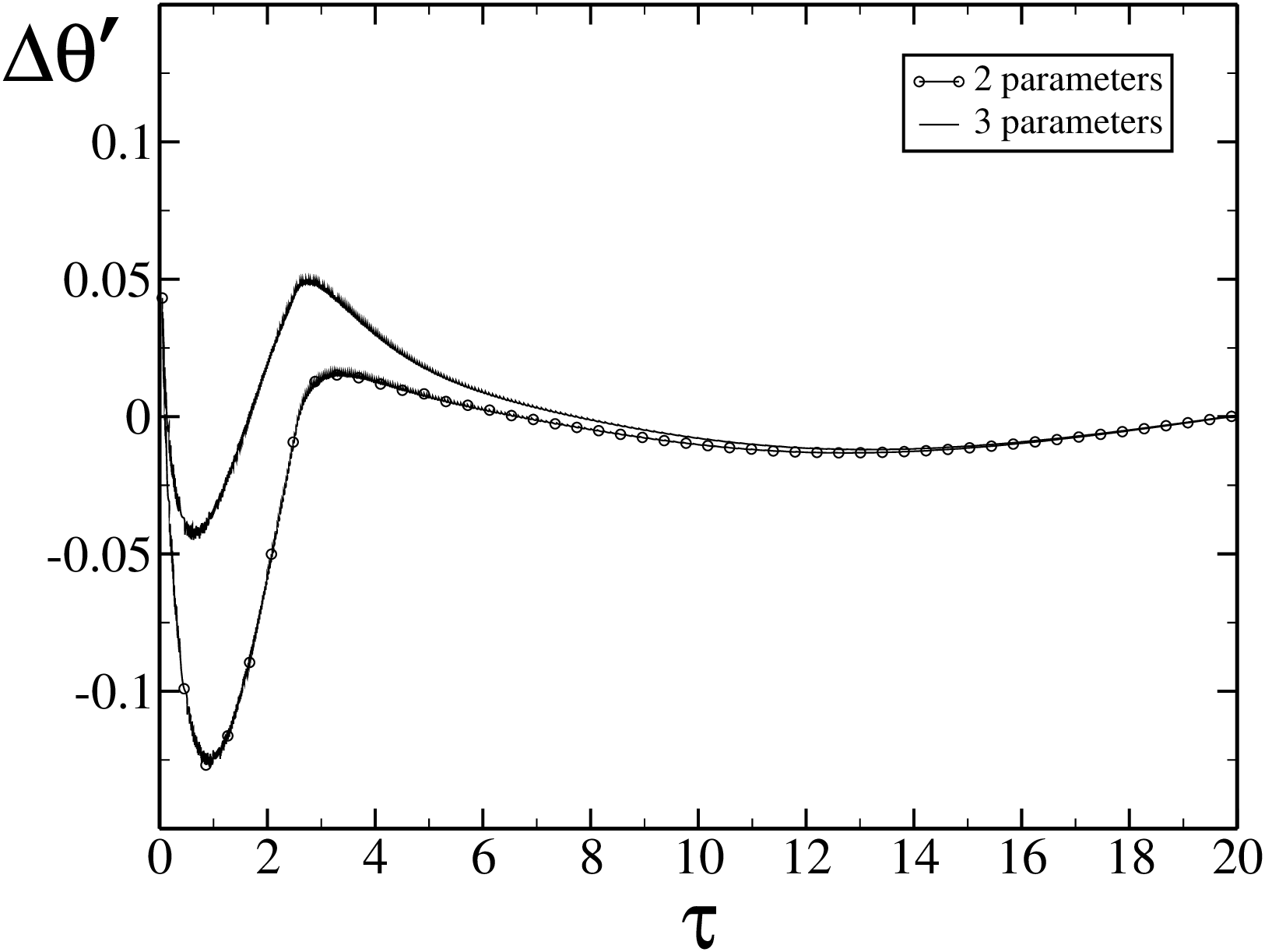} 
\caption{Jump-function $\Delta \Theta^{'} ( r^* )$ for two  and three  free 
parameters} 
} 
\end{figure} 
  
\section{The optimal fluctuation probability} 
 
The  solution of Eqs. (\ref{et}),(\ref{ep}) allows us to calculate the 
action  
\BA 
S&=&\int p \frac {\partial T} {\partial t} d^3 r dt -\int Hdt\nonumber \\  
\qquad &=&s_0\frac {n c_p{(T_{\infty}-T_c)}^2} {2{{T}^2}_{\infty}}{(\sqrt{\chi 
\tau_{ph}})}^3. 
\EA 
The negative constant $s_0$ is the dimensionless action  
\BE 
\label{dimlesact} 
s_0=\int \limits_{-\infty} \limits^{\infty } d\tau \int \partial_{\tau} \Theta \Pi 
d^3\xi  -\int \limits_{-\infty} \limits^{\infty } H_1 d\tau, 
\EE 
where the space integral is taken over the whole space. Here  
\BE 
\label{hamdl} 
H_1\!=\!\int  \Pi ({\bf r}) \left[   
{\bf \nabla}^2 \Pi ({\bf r})\!+\! {\bf  \nabla}^2\Theta \!+\! 
U(-\Theta) \Theta  \right] d^3\xi 
\EE 
is the dimensionless Hamiltonian, Eq.(\ref{ham}). 
The negative constant $s_0$ is a universal number  corresponding  
to the largest action $S_0$ and  is  
independent of the values of physical constants and the  
difference $T_{\infty}-T_c$.  
 
From Eq.(\ref{et}) and from the expression for the Hamiltonian, Eq. (\ref{hamdl}),  
it is easy to see that 
$$ 
s_0=\int \limits_{-\infty} \limits^{\infty } d\tau \int \Pi \nabla^2 \Pi d^3 r 
$$ 
for a smooth solution of Eqs.(\ref{et}), (\ref{ep}).  
In our case, the jump of $\Theta$ derivatives should be taken into account 
which modifies the action :  
$$ 
s_0=\int \limits_{-\infty} \limits^{\infty } d\tau \int \Pi \nabla^2 \Pi d^3 \xi -\int 
\limits_{-\infty} \limits^{\infty } d\tau \int \int \Delta \Theta^{'} \Pi dS_{r^*}.  
$$ 
The surface integral should be  taken over  
a sphere of radius $r^*$. Using  Eq. (\ref{ep}), 
$$ 
\nabla^2 \Pi = -\partial_{\tau} \Pi -U(r^*-\xi) \Pi, 
$$ 
  we finally 
find for the action 
\BE 
\label{modact} 
s_0=-\int \limits_{-\infty} \limits^{\infty } d\tau \int \limits_{\xi<r^* } \Pi^2 d^3 
\xi -\int \limits_{-\infty} \limits^{\infty } dt \int \int \Delta \Theta^{'} \Pi 
dS_{r^*}   
\EE 
Substituting the results of the numerical calculations we get 
$$ 
s_0=-100.73. 
$$ 
Test results for a smaller grid spacing, $\delta\xi=0.003$, give  
$$ 
s_0=-100.23 
$$ 
with a negligible  jump-function, $\Delta \Theta^{'}$.  
These results are in agreement with the naive estimate from  Eq.(\ref{modact}): 
The characteristic scale of $\Pi^2$ is 1 and the action is proportional to the volume of  
a sphere of radius $\pi$, which is about 100.

To estimate the temperature-variation rate, one can  
take  
\BE 
\label{maxvel} 
|v_{max}|=\frac {T_{\infty}-T_c} {\tau_{ph}}{(\sqrt{\chi \tau_{ph}})}^3 n. 
\EE 
In this case, in accordance with Eq. (\ref{2estim}), one can write  
for the probability flux in the transition region  
\BE 
\label{flux} 
j \sim \frac {T_{\infty} {(n {\left(\sqrt{\chi \tau_{ph}}\right)}^3)}^{\frac 1 
2}} {\tau_{ph} c_p} e^{S} \nu. 
\EE 
The constant $\nu$ cannot be estimated from the theory of  
hydrodynamic fluctuations \cite{PiT}. This quantity gives the  
number of small equilibrium fluctuations with $\delta T\ll T$   
in a unit volume on the atomic scale. As an estimate, we  
can use the relation $\nu={n}/T_{\infty}$ . Thus, the number of critical  
fluctuations  per unit time in a unit volume is  
\BE 
\label{numberinst} 
\frac {dN} {dt}\sim\frac{n {(n{{\left( \sqrt{\chi \tau_{ph} }\right)}^3)}^{ 1/2}}} 
 {\tau_{ph} c_p} 
\exp\left[s_0\frac {n c_p{(T_{\infty}-T_c)}^2} {2{{T}^2}_{\infty}} 
{(\sqrt{\chi \tau_{ph}})}^3\right]. 
\EE  
The nuclei of a new phase are intensively formed  
as $T_{\infty}$ approaches $T_c$ and then grow rapidly. We have  
considered the initial phase of critical-fluctuation  
growth and restricted our analysis  to the heat  
transfer via heat conduction, disregarding  superfluidity  
effects at this stage. This approximation can be  
justified by the fact that the largest contribution to the  
action comes from the region lying far from the region  
of low velocities $v$, where $p({\bf r})={\delta S}/{\delta T({\bf r})} $ becomes  
small and the fluctuation contribution can be neglected because $W\sim v^{-1}e^{S_0}$.  
 
\section{Bose condensation in the two-dimensional case} 
 
 Most   experiments  
on Bose-condensation in exciton systems were done in the two-dimensional (2D) case  
because these excitons are more stable \cite{But,Tim}.  
Condensation into the superfluid state of 
interwell exctions in AlAs/GaAs structures was supposedly observed in  
\cite{But}.  
In these experiments excitons were obtained in a 2D quantum well and were cooled 
by phonon emission into the 3D volume of the surrounding semiconductor \cite{SVI}. 
We can expect our theory to be suitable to 
explain the formation of a condensate in this case. 
In such systems the Bose-gas of excitons is dilute ($n {a_0}^2 \ll 1$) and has a long 
lifetime. There is no a true condensate at any nonzero 
temperature but it was shown in \cite{Pop} that there is a superfluid transition at  
\BE 
\label{tc2d} 
T_c= \frac {2 \pi n \hbar^2} {m \ln \ln( 1/{n {a_0}^2})}. 
\EE 
According to \cite{Pop} the pressure has the form: 
\BE 
\label{pres2d} 
P= \frac {2 \pi n^2}{m \ln( 1/ {n {a_0}^2})}+ \frac {m T^2 \zeta (2)} {2 
\pi \hbar^2}. 
\EE 
As it was pointed out in \cite{fish} these results hold only under the  
condition,  
\BE 
\label{lageparam} 
\ln \ln ( 1/ {n {a_0}^2}) \gg 1.  
\EE 
 Using the analogy to the  3D case, we consider temperature fluctuations 
at  constant pressure and generalize  
the results of  previous sections to the 2D case. 
In these  fluctuations below $T_c$  there is also an increase of the reciprocal phonon  
time, and in the 2D case  
 Eq.(\ref{nt}) should be replaced by 
\BE 
\label{denspres2d} 
\frac {\delta n } n =- \frac {\delta T} T  
\frac {\ln  (1 /(n {a_0}^2))} {{(\ln \ln  (1 /(n {a_0}^2)) 
)}^2}. 
\EE 
 The large parameter $ {(\ln  (1 /(n {a_0}^2))}/ {{(\ln \ln  (1 /(n {a_0}^2)) 
)}^2} $ plays the role of $1/{{\eta}^{\frac 1 3}}$ in 
Eq. (\ref{nt}). 
Eventually, we arrive to the  same set of equations, Eqs. (\ref{et}-\ref{ep}),  
as in the  3D case. 
Thus, the  number of critical fluctuations per unit time in a unit volume in the 
2D case is 
\BE 
\label{number2d} 
\frac {dN} {dt}\sim\frac{n {(n{{ \chi \tau_{ph} })}^{ 1/ 2}}} 
 {\tau_{ph} c_p} 
\exp\left[s_0\frac {n c_p{(T_{\infty}-T_c)}^2} {2{{T}^2}_{\infty}} 
{(\chi \tau_{ph})}\right ]. 
\EE 
 
Numerical calculations in the 2D case are  
somewhat more complicated in comparison to the 3D 
case. 
 There is no  a stationary solution like Eq.(\ref{statsol}).  
However, if we consider the system of a 
finite size $R$, there will be an analog to Eq.(\ref{statsol}): 
\BA 
\Theta_{stat}= -\frac {J_0 (\xi)} {{\lambda^1}_0 J_1 ( {\lambda^1}_0 ) \ln (\frac R 
{{\lambda^1}_0 })},~  \xi<{\lambda^1}_0, \nonumber \\ 
\Theta_{stat}= \frac {\ln (\frac \xi {\lambda^1}_0 ) } {\ln (\frac R {\lambda^1}_0 
)}, ~ \xi>{\lambda^1}_0.  
\EA 
Here $J_n$ is a Bessel function and ${\lambda^1}_0$ is the first zero of 
$J_0 (x)$. We see that the negative values of $\Theta$ are logarithmically suppressed. 
This is not very important because the system after passing in the vicinity of 
 the stationary 
point evolves further due to the divergence term  
(which was neglected in Eq.(\ref{ham}) and 
Eqs.(\ref{et}),(\ref{ep})). This  term in Eq. (\ref{ham}) has the form  
$$ 
\Delta H = \int \frac {U(T_c-T)}{\tau_{ph}} d^2 r. 
$$ 
The Hamilton equations, Eqs.(\ref{et}),(\ref{ep}), should  also be modified:  
\BE 
\frac {\partial\Theta} {\partial \tau}={\bf \nabla}^2 \Theta +\Theta 
U(-\Theta)+2{\bf \nabla}^2\Pi,   
\EE 
\BA 
\label{ep2} 
\frac {\partial\Pi} {\partial \tau}=&-&{\bf \nabla}^2 \Pi-\Pi 
U(-\Theta)+ \nonumber \\  
&&\delta (-\Theta) \frac {{T_{\infty}}^2}{{(T_{\infty}-T_c)}^2 n c_p \chi \tau_{ph}}. 
\EA 
Here $\delta (-\Theta)$ is a delta-function. As in the toy-model,  
 this term starts to play an important role at large $\tau$ while at small  
and intermediate $\tau$ 
it is unimportant. There is no need to take this term into account in our 
calculations of the action in the 3D case because it is a higher-order 
quasiclassical correction for the 3D instanton. But in the 2D case,   
we can consider the finite-time evolution  
 due to this term. The large-time cutoff can be estimated by assuming that at this  
moment  
the  divergence term becomes of the same order of magnitude  
as the main terms in Eq.(\ref{ep2}).  
We know 
the large-time asymptotics of  $\Pi$ in a finite-size  system of radius $R$,  
 $$ 
 \Pi \rightarrow \alpha \Psi_{\lambda} (r) \exp(- | \lambda | \tau), 
 $$ 
where $\Psi_{\lambda} (r)$ is the eigenfunction with the negative eigenvalue 
$\lambda$ of the 2D Schroedinger equation with the potential $-U({\lambda^1}_0 -\xi)$.  
Using this asym 
 
 ptotics and comparing terms in the equation for the norm, $|| \Pi ||= 
\int \Pi^2 d^2 r $,  
$$ 
\partial_{\tau} \frac {|| \Pi ||} {2} = - |\lambda| * || \Pi ||+\Pi 
({\lambda^1}_0) \frac {{T_{\infty}}^2}{{(T_{\infty}-T_c)}^2 n c_p \chi \tau_{ph}}, 
$$ 
we can estimate the large-time cutoff as: 
$$ 
\Delta \tau \sim \frac 1 {|\lambda|}  
\ln \left(\alpha |\lambda|\frac {{(T_{\infty}-T_c)}^2 n c_p  
\chi\tau_{ph}}{{T_{\infty}}^2}\right). 
$$ 
The space position of the thermal front, that corresponds to this time, is 
$$ 
\Delta r = \sqrt {\Delta \tau}. 
$$ 
If we choose the point $R$, where  $\Theta (R) =1$,  
at the distance $\Delta r$ from the point $r^*$ (where $\Theta ( r^* ) =0$ at 
the largest time), we can assume that the value of the total action will be almost 
independent of the exact position of $R$. We have performed a number of 
runs with different values of $R$ and have found no essential difference in the 
action. 
The resulting constant is  
$$ 
s_0=-13.6 
$$ 
The same naive estimate as in 3d can be performed. The action now is  
proportional $\pi 
{\left({{\lambda^1}_0}\right)}^2$ which is  about 13.

\section{The later stage of instanton growth} 
 
Analysis of the subsequent growth of the instanton  
requires the solution of hydrodynamic equations for a  
superfluid liquid, because a superfluid core appears in  
the developing fluctuation. We will qualitatively consider  
the phenomena that arise in this case. Proceeding  
from the assumption that the  value of $\tau_{ph}$ is high, we  
assume that the motion in this region is quasi-stationary  
 and tuned due to the slow cooling by phonons. We will use  
the hydrodynamic equations for a superfluid liquid in  
the vicinity of the transition point in the form proposed  
by Khalatnikov \cite{Hal}:  
$$ 
  \frac{\D {\bf v}_s}{\D t} = -\nabla \left( \frac{v_s^2}{2}+\mu+\mu_s \right), 
$$ 
$$ 
  \frac{\D\rho}{\D t} + \Div(\rho_s {\bf v}_s + \rho_n {\bf v}_n)=0, 
$$ 
$$ 
  \frac{\D}{\D t}   \left( \rho_s v_s^i + \rho_n v_n^i \right) + 
  \frac{\D}{\D x^k} \left( \rho_n v_n^i v_n^k + \rho_s v_s^i v_s^k + 
P\delta^{ik} \right)=0, 
$$ 
$$ 
  T\frac{\D (n\sigma)}{\D t} + T \Div (n\sigma {\bf v}_n)  
$$ 
$$   
  =\frac{2\Lambda m}{\hbar}{\left[ \mu_s+\frac{({\bf v}_n-{\bf v}_s)^2}{2} \right]}^2  
\rho_s-\frac{\rho_s c_p T}{\taup}, 
$$ 
$$ 
  \frac{\D\rho_s}{\D t} + \Div \rho_s {\bf v}_s = -\frac{2\Lambda m}{\hbar}{\left[ \mu_s+\frac{({\bf v}_n-{\bf v}_s)^2}{2} 
  \right]}\rho_s. 
$$ 
Here, $\sigma$ is the entropy per particle, n is the number of  
particles per unit volume, and $\rho$ is the density. The subscripts  
$n,s$ correspond to the normal and superfluid  
components, respectively; the constant $\Lambda$ is the relaxation  
parameter; and we introduced the term that  
accounts for the phonon-induced energy removal in the  
equation for entropy. Here, the specific chemical potential  
╣ $\mu_s$ for the superfluid density should ensure the condensate  
equilibrium density that is obtained by equating  
to zero the relaxation right-hand side of the equation  
for $\rho_s$. In our model of a weakly nonideal Bose gas,  
we can define phenomenologically  
\BE 
  \label{chem} 
  \mu_s = -\frac{\hbar^2 a_0}{m^2}\left[ (n-n(T))\right]+ 
   \frac{\hbar^2 a_0}{m^3}  
  \rho_s,  
\EE 
so that $\rho_s=m(n-n(T))=m \delta n$ in the equilibrium. Here,  
$n(T)$ is the number of particles outside the condensate.  
We assume that the quantity ${\Lambda m}/{\hbar}$ is large and $T_c-T$ is  
large enough for the approximate equality ╣$\mu_s+{v_s^2}/{2}\approx 0$ 
to be satisfied (we disregard quantity $v_n$, which is  
small compared to $v_s$); this gives  
\BE \label{Eq17} 
  \frac {\rho_s} m =\delta n - \frac{v_s^2 m^2}{2 \hbar^2 a_0} = 
  \delta n \left( 1 - \frac{v_s^2 m^2}{2 \hbar^2 a_0 \delta n} \right). 
\EE 
In this case, it follows from the hydrodynamic equations  
that $\mu\approx\mu(P,T)=const$,  
\BE 
  T\frac{(\D {n \sigma})}{\D t}+T\Div  
{(n \sigma {\bf v}_n)} = -\frac{c_p T\rho_s}{m \taup}, 
\EE 
and the momentum conservation law gives  
$$ 
  \frac{\D}{\D t}   \left( \rho_s v_s^i + \rho_n v_n^i \right) + 
  \frac{\D}{\D x^k} \left( \rho_n v_n^i v_n^k + \rho_s v_s^i v_s^k + 
  P\delta^{ik} \right)=0 
 $$ 
In view of the smallness of $v_s$  compared to the sound  
velocity and the smallness of $\rho_s$, we will neglect these  
corrections to pressure $P\approx P_0$. In this case, only the  
equation for entropy is significant. Assuming that the  
derivative ${(\D n\sigma)}/{\D t}$ is small, according to the assumption  
that the process is quasi-stationary (low temperature-variation rate) 
, we find that the 
stationary regime $-\Div(\sigma n {\bf v_n})=-{c_p\rho_s}/{\taup}$ should approximately  
take place and that the mass flux should be zero  
($\rho_s v_s+\rho_n v_n=0 $). Considering that $\rho_n\approx\rho$ we obtain the  
equation  
\BE 
  -\sigma \Div {(\rho_s v_s)} = -\frac{c_p\rho_s}{\taup} 
\EE 
where $\sigma$ is the entropy per particle. This equation determines  
the heat transfer in the fluctuation superfluid  
core. Using Eq.(\ref{Eq17}) we obtain  
$$ 
  \sigma \frac{1}{r^2}\frac{\D}{\D r}r^2(1-\frac{v_s^2}{u^2})v_s - 
  (1-\frac{v_s^2}{u^2})\frac{c_p}{\taup}=0, 
$$ 
\BE 
   u^2=\frac{2 \hbar^2 a_0 \delta n}{m^2}. 
\EE 
By introducing the dimensional distance $\xi=c_p r/\sigma u \taup$ 
and $v=v_s/ u$ 
we arrive at the equation  
 
\BE 
   \frac{\D v}{\D \xi} = \frac{(1-v^2)(1-\frac{2}{\xi}v)}{1-3v^2} 
\EEб√

The singular points of this differential equation are  
$$ 
  \xi=0,~ v=0  ~~~\mbox{and}~~~  \xi=\frac{2}{\sqrt{3}},~ v=\frac{1}{\sqrt{3}}, 
$$  
the latter point being a focus with the eigenvalues $\lambda=1\pm i\sqrt{5}$ 
 Since the velocity $v$ must vanish at $\xi=0$, $v\approx\xi/3$  
for small $\xi$ and increases faster than by the linear 
law, with the derivative $\frac{d\xi}{d v}$ vanishing at $v=\frac{1}{\sqrt{3}}$  
and at a certain $\xi=\xi_c$, whereupon the derivatives  
assume negative values upon the further increase in $v$.  
Thus, a regular superfluid flow cannot be continued  
after the point $\xi_c$ (the constant is on the order of unity  
and can be determined numerically).  
The critical radius  
\BE 
  r_c\sim \frac{\sigma u \taup}{c_p} = 
  \frac \sigma {c_p} \sqrt{2\frac{\delta n}{n}\eta^{1/3}\frac{\taup}{\tau_{tr}}}\sqrt{\chi\taup} 
\EE 
can be smaller than $\sqrt{\chi \tau_{ph}} $ ; it should also be noted that  
$1-v^2>0 $(i.e., a singularity appears in the superfluid core).  
This singularity indicates that the quasi-stationarity  
conditions are violated at $\xi\gtrsim \xi_c $, and a complex  
nonstationary superfluid flow with the intense vortex  
formation in an instanton should appear upon the transition  
to the normal liquid at $T > T_c$. Similar effects are  
observed in a superfluid liquid in the gravitational field,  
where $T_c$ is a function of one (vertical) coordinate and a  
fixed heat flux from the superfluid to the normal liquid  
takes place \cite{Ak}. We are dealing with a similar situation  
arising due to the nonuniform cooling as the critical  
temperature in the superfluid nucleus is approached.  
The results of numerical calculation \cite{Wen} and experimental  
data \cite{Fen},\cite{Bad} indicate the formation of a ''vortex'' 
superfluid phase with a higher but finite thermal  
conductivity without a superfluid transport. The mechanism  
of vortex formation and the vortex phase of this  
kind have been poorly studied both theoretically and  
experimentally.  
 
\section{conclusion.} 
Thus, we have shown that, in contrast to \cite{Zur}, a transition  
to the superfluid phase can occur through an independent  
growth of critical fluctuations (instantons) at  
temperatures above the critical point $(T > T_c) $ immediately  
in the course of external cooling. These fluctuations  
subsequently transform into macroscopic formations.  
The growth of the nucleus of the superfluid state  
is accompanied by vortex generation in its external  
part. Consequently, vortex defects appear both due to  
the independent nucleation with an arbitrary phase  
upon cooling (the Zeldovich-Kibble hypothesis) and  
directly during the growth of each superfluid nucleus.  
This vortex-generation mechanism during the growth  
of an instanton significantly differs from the mechanism  
determined in \cite{Ar}, where the existence of a  
superfluid flow interacting with the heated normal  
regions was presumed. In \cite{Ar}, an attempt was made to  
explain the results of experiments \cite{Ru}, in which $^3 \rm{He}$  
was irradiated by neutrons. As a result, some regions heated  
to temperatures above $T_c$ appeared. These regions were  
cooled by the surrounding superfluid $^3 \rm{He}$, and the formation  
of vortices was detected. Thus, nonuniform  
cooling took place that differs considerably from the  
model used in our study. In the critical fluctuation considered  
here, heating takes place due to its nonsuperfluid  
surroundings. Consequently, it is advantageous  
for the fluctuation to preserve its spherical symmetry to  
reduce this heating. In the case of cooling of a heated  
region with superfluid surroundings \cite{Ru}, the interface  
must obviously be unstable against its shape distortions,  
because this leads to a faster cooling. However,  
the stability, as well as the phase-transition mechanism  
itself, under such conditions (which, in contrast to \cite{Ar},  
are not associated with the existence of an external  
superfluid flow) calls for detailed investigations.  
 
Probably, analogous schemes can be developed for the kinetics of various  
other phase transitions in the presence of the external cooling with  
the scenario essentially given by the toy model described in the text. 
 
\section{ACKNOWLEDGMENTS}  
We are grateful to V.V. Lebedev, I.V. Kolokolov and H. M\"uller-Krumbhaar 
for discussions.  
This study was supported by the President of the  
Russian Federation (grant no. NSh-1715.2003.3) in  
Support of Young Russian Scientists and Leading Scientific  
Schools, the program б⌠Quantum Macrophysicsб■  
of the Presidium of the Russian Academy of Sciences,  and RFFI  
grants no. 03-02-16012, 05-02-16553.


\newpage 
\begin{thebibliography}{99} 
\bibitem{Bec} R.Becker, W.Doering. Annalen der Physik  {\bf 24} , 719 (1935) 
\bibitem{Zel} Ya. B. Zel'dovich, Zh. Exp. Teor. Fiz. {\bf 112}, 525 (1942) 
\bibitem{Lan} J.S. Langer. Ann. of Physics {\bf 54} 258 (1962) 
\bibitem{Lif} I. M. Lifshits, Zh. Exp. Teor. Fiz {\bf 42},  1354 (1962) 
[Sov. Phys. JETP {\bf 15}, 939 (1962)] 
\bibitem{Zell} Ya. B. Zel'dovich, I. Yu. Kobzarev, and L. B. Okun', Zh.  
Exp. Teor. Fiz. {\bf 67} , 3, (1974) 
[Sov. Phys. JETP {\bf 40}, 1 (1975)] 
\bibitem{Kib} T.W Kibble, J.Phys. {\bf A9}, 1387 (1976) 
\bibitem{Zur} W.H. Zurek, "Cosmological experiments in condensed matter system", 
 Physics Reports 177-221 (1996) 
\bibitem{BIS} E.A. Brener, S.V. Iordanskiy, R.B. Saptsov, JETP Letters {\bf 79}, 
410 (2004)  
\bibitem{Henry} H. Henry and Herbert Levine, PRE {\bf 68}, 031914 (2003) 
\bibitem{Marder} M. Marder, PRE {\bf 54}, 3442 (1996) 
\bibitem{Cher} A.I. Chernykh, M.G. Stepanov, PRE {\bf 64}, 026306 (2001) 
\bibitem{Ld} L. D. Landau and E. M. Lifshitz, Course of Theoretical  
Physics, Vol.{\bf 5} "Statistical Physics", Fizmatlit,  
Moscow, 2002; Pergamon Press, Oxford, 1980. 
\bibitem{Liff}E. M. Lifshitz and L. P. Pitaevskii, Course of Theoretical  
Physics, Vol.{\bf 9} Statistical Physics  p.2. h. 9, Fizmatlit,  
Moscow(2002); Pergamon, New York, 1980 
\bibitem{Kl}  V. I. Klyatskin, Stochastic Equations by the Eyes of a  
Physicist, Fizmatlit,  
Moscow (2001) 
\bibitem{PiT} E. M. Lifshitz and L. P. Pitaevskii, Physical Kinetics  
(Nauka, Moscow, 1979; Pergamon Press, Oxford, 1981).  
\bibitem{But} L.V. Butov, A.I. Filin. PRB {\bf 58}, 1980 (1998) 
\bibitem{Tim} A.V. Larionov et. al. JETP Letters {\bf 75}, 570 
(2002) 
\bibitem{SVI} S.V. Iordanskiy, A.B. Kashuba. JETP Letters {\bf 73}, 542-545 
(2001) 
\bibitem{Pop} V.N. Popov , Functional Integrals in Quantum Field Theory and 
Statistical Physics, Atomizdat, Moscow (1976). (Reidel,Dordrecht,1983). 
\bibitem{fish} D.S. Fisher, P.C. Hohenberg. PRB {\bf 37}, 4936 (1988) 
 
\bibitem{Hal} I. M. Khalatnikov, The Theory of Superfluidity (Nauka,  
Moscow, 1971), Chap. 9 
\bibitem{Ak} Akira Onuki. J. Low Temp. Physics {\bf 50},  5/6, 433 (1982) 
\bibitem{Wen} P.B. Weinman, J. Miller. J. Low Temp. Physics {\bf 119}, 155 (2000) 
\bibitem{Fen} Feng Chuan Lui, Guenter Ahlers. PRL, {\bf 76},  8, 1300 (1996) 
\bibitem{Bad} H. Baddar, G. Ahlers, Kuehn, H. Fu.  J. Low Temp. Physics {\bf 
119},  1/2, 1 (2000) 
\bibitem{Ar} I.S. Aranson, N.B. Kopnin, V.M. Vinokur. PRL {\bf 83}, 2000 (1999) 
\bibitem{Ru} V.H.M Ruutu, V.B Eltsov, A.J Gill et al., Nature {\bf 382}, 334 (1996) 
 
 
\end{thebibliography}
\end{document}